\documentclass[conference]{IEEEtran}
\ifCLASSINFOpdf
\else
\fi
\usepackage{amsmath,amssymb,amsfonts,amsthm}
\DeclareMathOperator*{\argmin}{arg\,min}
\usepackage{mathtools}
\usepackage{color}
{
    \theoremstyle{plain}
    \newtheorem{assumption}{Assumption}
}
\begin{document}
%
\title{An Implementation of the Poisson Multi-Bernoulli Mixture Trajectory Filter via Dual Decomposition}
%
%
%
\author{\IEEEauthorblockN{Yuxuan Xia\IEEEauthorrefmark{2},
Karl Granstr\"{o}m\IEEEauthorrefmark{2},
Lennart Svensson\IEEEauthorrefmark{2}, 
\'{A}ngel F. Garc\'{i}a-Fern\'{a}ndez\IEEEauthorrefmark{3}}
\IEEEauthorblockA{\IEEEauthorrefmark{2}Department of Electrical Engineering, Chalmers University of Technology, G\"{o}teborg, Sweden}
\IEEEauthorblockA{\IEEEauthorrefmark{3}Department of Electrical Engineering and Electronics, University of Liverpool,
Liverpool, United Kingdom}
}

\maketitle

\begin{abstract}
This paper proposes an efficient implementation of the Poisson multi-Bernoulli mixture (PMBM) trajectory filter. The proposed implementation performs track-oriented N-scan pruning to limit complexity, and uses dual decomposition to solve the involved multi-frame assignment problem. In contrast to the existing PMBM filter for sets of targets, the PMBM trajectory filter is based on sets of trajectories which ensures that track continuity is formally maintained. The resulting filter is an efficient and scalable approximation to a Bayes optimal multi-target tracking algorithm, and its performance is compared, in a simulation study, to the PMBM target filter, and the delta generalized labelled multi-Bernoulli filter, in terms of state/trajectory estimation error and computational time.
\end{abstract}



%
\IEEEpeerreviewmaketitle

\section{Introduction}
Multiple target tracking (MTT) refers to the problem of jointly estimating the number of targets and their trajectories from noisy sensor measurements \cite{bar1990multitarget}. The three most popular approaches to solving the MTT are the joint probabilistic data association filter \cite{jpda}, the multiple hypothesis tracker (MHT) \cite{homht,tomht,blackman2004multiple} and multi-target filters based on random finite sets (RFS) \cite{rfs}. 


During the last decade, RFS-based MTT algorithms have received a great deal of attention. In the early stage, RFS-based MTT algorithms were developed based on moment approximations of posterior multi-target densities, including the Probability Hypothesis Density (PHD) filter \cite{phd1} and the Cardinalised PHD (CPHD) filter \cite{cphd1}. In recent years, a significant trend in RFS-based MTT is the development of conjugate multi-target distributions \cite{glmbconjugateprior,pmbmpoint}, meaning that the posterior multi-target distribution has the same functional form as the prior. Comparisons have shown that PHD and CPHD filters have worse tracking performance, compared to filters based on MTT conjugate priors \cite{glmbpoint,pmbmpoint2}.

We proceed to describe two MTT conjugate priors in the literature for the standard point target measurement model, which assumes that each target gives rise to at most one measurement and that targets do not have any shared measurement. The first conjugate prior, presented for unlabelled RFSs in \cite{pmbmpoint}, consists of the disjoint union of a Poisson point process (PPP) describing targets that have not yet been detected, and a multi-Bernoulli mixture (MBM) considering all the data association hypotheses. The second conjugate prior is the generalized labelled multi-Bernoulli (GLMB) density \cite{glmbconjugateprior} presented for labelled RFS, in which labels are appended to target states with the objective of forming target trajectories. 

The Poisson multi-Bernoulli mixture (PMBM) filter \cite{pmbmpoint2} and the $\delta$-GLMB filter \cite{glmbpoint} are two computationally tractable approaches, respectively, based on the PMBM and GLMB conjugate priors. The PMBM filter has computational advantages in relation to the $\delta$-GLMB filter due to the structure of the hypotheses \cite{pmbmpoint2}. This was further analyzed in the performance evaluation study for point target filtering in \cite{performanceevaluation}. However, unlike the $\delta$-GLMB filter, the PMBM filter on its own does not provide information to estimate trajectories.


A Bayesian formulation of the MTT problem, in terms of unlabelled RFS of trajectories, was recently provided in \cite{trackingbasedontrajectories}. The PMBM filtering recursions based on this framework have been derived in \cite{continuityPMBM}, which enables us to leverage on the benefits with the PMBM recursions while also obtaining track continuity. As a sequel to \cite{continuityPMBM}, the aim of this paper is to complement its theoretical contributions with practical algorithms. In particular, we present an efficient implementation of the PMBM trajectory filter that estimates the trajectories of the targets which are present in the surveillance area at the current time.

In this paper, we propose an approximation to the exact PMBM trajectory filter, by using track-oriented N-scan pruning \cite{blackman2004multiple} and dual decomposition \cite{dualdecomposition} to solve the multi-frame assignment problem. The proposed algorithm therefore shares some of the key properties of certain MHT algorithms \cite{blackman2004multiple,dualdecomposition}, but is derived using RFS and birth/death models. Similar work has been done in \cite{mori2016three}, where MHT algorithms considering multi-frame data association based on random finite set/sequence formalism are derived but with the assumption that the number of targets is unknown but constant over time. 

Compared to the $\delta$-GLMB filter, our algorithm is unlabelled, and, given a global hypothesis, we obtain a distribution over the set of trajectories based on which it is convenient to construct trajectory estimates. The performance of the PMBM trajectory filter and the existing PMBM target filter \cite{pmbmpoint2}, along with the $\delta$-GLMB filter with joint prediction and update step \cite{gibbsglmb}, are compared in a simulated scenario with coalescence.

The paper is organized as follows. In Section II, we introduce the background on RFS modelling and the PMBM conjugate prior on the set of trajectories. In Section III, we present our proposed implementation. Simulation results are presented in Section IV, and conclusions are drawn in Section~V. 

\section{Background}

In this section, we first outline the modelling assumptions utilized in this work. Next, we give a brief introduction to RFS of trajectories. Then, we give a summary to the PMBM conjugate prior on the set of trajectories.

\subsection{Modelling assumptions}

In the traditional RFS of targets formulation, target states and measurements are represented in the form of finite sets \cite{rfs}. Let $x_k$ denote a single target state at discrete time step $k$, and let $z^j_k$ denote the $j$th measurement at discrete time step $k$. The set of measurements obtained at time step $k$ is denoted as $Z_k$, including clutter and target-generated measurements with unknown origin. 

In this paper, the state variable is a RFS of trajectories, as suggested in \cite{trackingbasedontrajectories}. Single trajectories are parameterized using a tuple $X = (\beta,\epsilon,\mathbf{x}_{\beta:\epsilon})$ \cite{continuityPMBM,svensson2014target}, where $\beta$ is the discrete time of the trajectory birth, i.e., the time the trajectory begins; $\epsilon$ is the discrete time of the trajectory's most recent state (If $k$ is the current time, $\epsilon<k$ means that the trajectory ends at time $\epsilon$, and $\epsilon=k$ means that the trajectory is ongoing); $\mathbf{x}_{\beta:\epsilon}$ is, given $\beta$ and $\epsilon$, the sequence of states $(x_\beta, x_{\beta+1},...,x_{\epsilon-1},x_{\epsilon})$.

We utilize the standard multi-target dynamics model, defined in Assumption 1, and the standard point target measurement model, defined in Assumption 2.

\begin{assumption}
The multiple target state evolves according to the following time dynamics process:
\begin{itemize}
    \item Targets arrive at each time according to a PPP with intensity $\lambda^b(x_k)$, independently of any existing targets.
    \item Targets survive with a probability $p^s(x_k)$, and targets depart the surveillance area according to i.i.d. Markov processes with probability $1-p^s(x_k)$.
    \item Target motion follows i.i.d. Markov processes with transition density $\pi(x_{k}|x_{k-1})$.
\end{itemize}
\end{assumption}

\begin{assumption}
The multiple target measurement process is as follows:
\begin{itemize}
    \item Each target can generate at most one measurement per time scan with detection probability $p^d(x_k)$.
    \item The clutter measurements are modelled a PPP with Poisson rate $\lambda^c(z_k)$ and spatial distribution $c(z_k)$, and the clutter PPP intensity is $\kappa(z_k)$. Clutter measurements are independently of any targets and target-generated measurements.
    \item Each target-generated measurement is only conditioned on its corresponding target. The single target measurement likelihood is $\ell(z_k|x_k)$.
\end{itemize}
\end{assumption}


Expressions for recursively predicting and updating the multi-trajectory filtering densities based on the above modelling assumptions can be found in \cite{trackingbasedontrajectories}. Following \cite{continuityPMBM}, we recursively express these densities as PMBMs.




\subsection{Densities of sets of trajectories}

The basic building blocks in a PMBM density are the PPP and the Bernoulli process. Let $\mathbf{X}$ denote the set of trajectories, and let $f(\mathbf{X})$ denote the corresponding density function. A trajectory PPP is analogous to a target PPP, and has set density
\begin{equation}
    f^{\textrm{ppp}}(\mathbf{X}) = e^{-\langle\lambda^u,1\rangle}\prod_{X\in\mathbf{X}}\lambda(X),
\end{equation}
where $\langle a,b\rangle$ denotes $\int a(x)b(x)dx$. The trajectory PPP intensity $\lambda(\cdot)$ is defined on the trajectory state space, i.e., realizations of the PPP are trajectories with a birth time, a time of the most recent state, and a trajectory. 

A trajectory Bernoulli process is analogous to a target Bernoulli, and has set density
\begin{equation}
f^{\textrm{ber}}(\mathbf{X}) = \begin{cases}
    1-r& \mathbf{X}=\emptyset\\
    r\cdot f(X)& \mathbf{X}=\{X\}\\
    0& \text{otherwise},
\end{cases}
\end{equation}
where $f(X)$ is a trajectory state density, and $r$ is the Bernoulli probability of existence. Together, $r$ and $f(\cdot)$ can be used to find the probability that the target trajectory existed at a specific time, or find the probability that the target state was in a certain area at a certain time. Trajectory multi-Bernoulli (MB) RFS and trajectory MBM RFS are both defined analogously to target MB RFS and target MBM RFS: a trajectory MB is the disjoint union of multiple trajectory Bernoulli RFS; trajectory MBM RFS is an RFS whose density is a mixture of trajectory MB densities.

\subsection{PMBM trajectory filter}

\subsubsection{PMBM conjugate prior on the set of trajectories}

The PMBM conjugate prior was developed for point targets in \cite{pmbmpoint} and extended to trajectories in \cite{continuityPMBM}. In this context, conjugacy means that the family of PMBM densities is closed under the prediction and update steps, though the number of terms in the mixture grows rapidly with time. In the trajectory PMBM form, the trajectory set $\mathbf{X}$ is a union of two disjoint sets $\mathbf{X}^u$ and $\mathbf{X}^d$, i.e., $\mathbf{X}=\mathbf{X}^u\uplus\mathbf{X}^d$. The trajectories in the set $\mathbf{X}^u$ are hypothesised to exist but to be undetected, and $\mathbf{X}^u$ is presented by a PPP. The set of detected trajectories $\mathbf{X}^d$ is presented by an MBM. The PMBM density on the set of trajectories can be expressed as
\begin{subequations}
\begin{align}
    f(\mathbf{X}) &= \sum_{\mathbf{X}^u\uplus\mathbf{X}^d=\mathbf{X}}f^{\textrm{ppp}}(\mathbf{X}^u)\sum_{j\in\mathbb{J}}W^jf^j(\mathbf{X}^d),\\
    f^{\textrm{ppp}}(\mathbf{X}^u) &= e^{-\langle \lambda^u,1\rangle}\prod_{X\in\mathbf{X}^u}\lambda^u(X),\\
    f^j(\mathbf{X}^d) &= \sum_{\uplus_{i\in\mathbb{I}^j}\mathbf{X}^i=\mathbf{X}^d}\prod_{i\in\mathbb{I}^j}f^{j,i}(\mathbf{X}^i),
    \label{eq:mbmdensity}
\end{align}
\label{eq:pmbm}%
\end{subequations}
where $f^{j,i}(\mathbf{X}^i)$ denotes the $i$th Bernoulli component in the $j$th MB $f^j(\mathbf{X}^d)$ of the MBM, and $W^j$ is the weight of the $j$th MB.


\subsubsection{Structure of the hypotheses}
The structure of the hypotheses in the PMBM trajectory filter is the same as the structure of the hypotheses in the PMBM target filter \cite{pmbmpoint}. In the PMBM trajectory filter, each MB corresponds to a unique \textit{global hypothesis} for the detected trajectories. A global hypothesis is a partitioning of all measurements received so far into subsets, where each subset is hypothesised to correspond to a \textit{single trajectory hypothesis}. A \textit{single trajectory hypothesis} is a subset of measurements, at most one from each time scan, that is hypothesised to correspond to the same target. To be consistent with the terminology used in \cite{pmbmpoint}, a \textit{track} is defined as a collection of single trajectory hypotheses, representing different possibilities of measurement sequences corresponding to the trajectory. We initiate one new track per measurement, and all single trajectory hypotheses initiated by the same measurement belong to the same track. Each global hypothesis must incorporate one single trajectory hypothesis from each track. 






Let $m_k$ be the number of measurements at time $k\in\{1,...,\tau\}$ and $j\in\mathbb{M}_k=\{1,...,m_k\}$ be an index to each measurement with value $z_k^j\in Z_k$. Let $\mathcal{M}_{k^{\prime}}$ denote the set of all measurement indices up to and including time step $k^{\prime}$; the elements of $\mathcal{M}_{k^{\prime}}$ are of the form $(\tau,j)$, where $j\in\{1,...,m_{\tau}\}$ is an index of a measurement in scan $\tau\leq k^{\prime}$. Let the number of tracks at time $k$ be $n_{k|k^{\prime}}$ and $i\in\mathbb{X}_{k|k^{\prime}}=\{1,...,n_{k|k^{\prime}}\}$ be an index to each track, where $k^{\prime}=k-1$ corresponds to the prediction step and $k^{\prime}=k$ corresponds to the update step. A global hypothesis at time $k$ can be represented as $A_{k|k^{\prime}}=(\mathbf{a}_{k|k^{\prime}}^1,...,\mathbf{a}_{k|k^{\prime}}^{n_{k|k^{\prime}}})$, where $\mathbf{a}_{k|k^{\prime}}^i$ denotes the single trajectory hypothesis utilized for the $i$th track. The set of single trajectory hypotheses for the $i$th track at time $k$ is denoted as $\mathcal{A}_{k|k^{\prime}}^i$. The set of measurement indices under single trajectory hypothesis $\mathbf{a}_{k|k^{\prime}}^i$ is represented as $\mathcal{M}(\mathbf{a}^i_{k|k^{\prime}})\subseteq\mathcal{M}_{k^{\prime}}$. 

In the PMBM trajectory filter, each global hypothesis should explain the association of each measurement received so far. In addition, every measurement should be associated to one and only one (pre-existing or new) track in each global hypothesis. In other words, the single trajectory hypotheses included in a given global hypothesis cannot have any shared measurement. Under these constraints, the set of global hypotheses at time scan $k$ can be expressed as \cite{pmbmpoint}
\begin{multline}
    \mathfrak{A}_{k|k^{\prime}}=\bigg\{(\mathbf{a}^1_{k|k^{\prime}},...,\mathbf{a}^{n_{k|k^{\prime}}}_{k|k^{\prime}})\bigg|\mathbf{a}^i_{k|k^{\prime}}\in\mathcal{A}^i_{k|k^{\prime}}, \bigcup_{i\in\mathbb{X}_{k|k^{\prime}}}\mathcal{M}(\mathbf{a}^i_{k|k^{\prime}})\\=\mathcal{M}_{k^{\prime}}, \mathcal{M}(\mathbf{a}^i_{k|k^{\prime}})\cap\mathcal{M}(\mathbf{a}^j_{k|k^{\prime}})=\emptyset~\forall~i\neq j\bigg\}.
    \label{eq:globalhypothesis}
\end{multline}

\subsubsection{Filtering recursion}

The form of the PMBM conjugate prior on the sets of trajectories is maintained by prediction and update steps. Given the sequences of measurements up to time step $k$, our objective is to calculate the PMBM multi-trajectory filtering density at time $k$ recursively. In the prediction step, the MBs describing pre-existing trajectories and the PPP describing undetected trajectories are predicted independently. By using a PPP birth model, the density of new born trajectories can be easily incorporated into the predicted density of undetected trajectories \cite{pmbmpoint}. 
In the update step, the PPP and the MBs are updated independently. The PPP intensity of undetected trajectories is thinned by the missed detection probability, i.e., $1-p^d(x_k)$. According to the point target PMBM modelling assumptions, each measurement creates a new track; thus, the number of tracks after updating becomes $n_{k|k}=n_{k|k-1}+m_k$. 

In what follows, we only present part of the update equations that are important to explain our proposed implementation due to page limits. We refer the reader to \cite{continuityPMBM} for more details. 

Let $w^{\mathbf{a}^i}_{k|k^{\prime}}$ denote the weight of single trajectory hypothesis $\mathbf{a}^i_{k|k^{\prime}}$, and let the corresponding Bernoulli density denote as $f^{\mathbf{a}^i}_{k|k^{\prime}}(\mathbf{X})$. For pre-existing tracks ($i\in\{1,...,n_{k|k-1}\}$), each single trajectory hypothesis creates $1+m_k$ new single trajectory hypotheses, one for missed detection,
\begin{subequations}
\begin{align}
    \mathcal{M}_k(\mathbf{a}^i)&=\mathcal{M}_{k-1}(\mathbf{a}^i),\\
    w^{\mathbf{a}^i}_{k|k}&=w^{\mathbf{a}^i}_{k|k-1}\mathcal{L}_k^{\emptyset,\mathbf{a}^i},\\
    \mathcal{L}_k^{\emptyset,\mathbf{a}^i} &= (1-r^{\mathbf{a}^i}_{k|k-1}+r^{\mathbf{a}^i}_{k|k-1}\langle f^{\mathbf{a}^i}_{k|k-1},1-p^d\rangle),
\end{align}
\end{subequations}
where $\mathcal{L}_k^{\emptyset,\mathbf{a}^i}$ is the likelihood that the potential trajectory represented by $\mathbf{a}^i$ is missed detected, and the others for measurement updates ($j\in\mathbb{M}_k$), 
\begin{subequations}
\begin{align}
    \mathcal{M}_k(\mathbf{a}^i)&=\mathcal{M}_{k-1}(\mathbf{a}^i)\cup\{(k,j)\},\\
    w^{\mathbf{a}^i}_{k|k}&=w^{\mathbf{a}^i}_{k|k-1}\mathcal{L}_k^{j,\mathbf{a}^i},\\
    \mathcal{L}_k^{j,\mathbf{a}^i}&=r^{\mathbf{a}^i}_{k|k-1}\langle f^{\mathbf{a}^i}_{k|k-1},\ell(z^j_k|\cdot)p^d\rangle,
\end{align}
\end{subequations}
where $\mathcal{L}_k^{j,\mathbf{a}^i}$ is the likelihood that single trajectory hypothesis $\mathbf{a}^i$ is updated by the $j$th measurement from time $k$.

Each new track ($i\in\{n_{k|k-1}+1,...,n_{k|k-1}+m_k\}$) contains two single trajectory hypotheses. The first one corresponds to a Bernoulli density with zero existence probability, and it covers the case that the new measurement is associated with one of the pre-existing tracks, 
\begin{subequations}
\begin{align}
    \mathcal{M}_k(\mathbf{a}^i)&=\emptyset,\\
    w^{\mathbf{a}^i}_{k|k}&=1.
\end{align}
\end{subequations}
The second one results from updating the PPP of undetected trajectories with a new measurement, which can be either a false alarm or the first detection of an undetected trajectory ($j\in\mathbb{M}_k$),
\begin{subequations}
\begin{align}
    \mathcal{M}_{k}(\mathbf{a}^i)&=\{(k,j)\},\\
    w^{\mathbf{a}^i}_{k|k}&=\mathcal{L}_k^{u,j,\mathbf{a}^i}= \kappa(z^j_k)+\langle D^u_k,\ell(z^j_k|\cdot)p^d\rangle,
    \label{eq:assolikelihoodnewdetection}
\end{align}
\end{subequations}
where $\mathcal{L}_k^{u,j,\mathbf{a}^i}$ is the likelihood that the $j$th measurement from time $k$ is assigned to an undetected trajectory.

\section{Implementation of the PMBM Trajectory Filter via Multi-frame Assignment}

In this section, we present an efficient implementation of the PMBM trajectory filter that estimates the trajectories of the targets who are present in the surveillance area at the current time. The proposed implementation uses track-oriented $N$-scan pruning to reduce the number of global hypotheses, and dual decomposition to obtain the most likely global hypothesis by solving a multi-frame assignment problem. In addition, we discuss the connections and differences between the existing PMBM target filter and the proposed PMBM trajectory filter.




\subsection{Global hypothesis reduction}
In the update step of the PMBM trajectory filter, each possible data association will create an updated global hypothesis so that the number of MBs in the PMBM posterior density (\ref{eq:pmbm}) will increase exponentially over time. Hence, finding a feasible approach to reduce the number of global hypotheses after the update step is essential for designing a computationally tractable PMBM trajectory filter. 

In the PMBM trajectory filter, the posterior over the set of detected trajectories is an MBM, i.e., a weighted mixture of global hypotheses. Typically, the estimates of trajectories are extracted from the most likely global hypothesis. Conditioning on the most likely global hypothesis, we make use of the track-oriented $N$-scan pruning \cite{blackman2004multiple}, a conventional hypothesis reduction technique used in MHT, to prune global hypotheses with negligible weights. 

Given the most likely global hypothesis $A^{*}$ at current time scan $\tau$, we trace the single trajectory hypotheses included in $A^{*}$ back to their local hypotheses at scan $\tau-N$. The assumption behind the $N$-scan pruning method is that the data association ambiguity is resolved before scan $\tau-N$ \cite{blackman2004multiple}. In other words, global hypotheses which do not have the same local hypotheses at scan $\tau-N+1$ as $A^{*}$ are assumed to have negligible weights; these global hypotheses can then be pruned. At last, tracks ending up with only non-existence single trajectory hypothesis can be pruned. In what follows, we show that the most likely global hypothesis $A^{*}$ can be obtained as the solution of a multi-frame assignment problem.

\subsection{Data association modelling and problem formulation}

The posterior global hypothesis weight $W^A_{k|k}$ is proportional to the product of different weights of the single trajectory hypotheses $w^{\mathbf{a}^i_{k|k}}$, one from each track \cite{pmbmpoint}:
\begin{equation}
    W^A_{k|k}\propto\prod_{i\in\mathbb{X}_{k|k}}w^{\mathbf{a}^i_{k|k}},
    \label{eq:globalhypothesisweight}
\end{equation}
where the proportionality denotes that normalization is required to ensure that $\sum_{A_{k|k}\in\mathfrak{A}_{k|k}}W^{A}_{k|k}=1$. Omitting time indices and introducing the notation $c^A=-\log (W^A)$ and $c(\mathbf{a}^i)=-\log (w^{\mathbf{a}^i})$, yields
\begin{equation}
    c^A = \sum_{i\in\mathbb{X}}c(\mathbf{a}^i) + C,
\end{equation}
where $C$ is the logarithm of the normalization constant in (\ref{eq:globalhypothesisweight}). The most likely global hypothesis is the collection of single trajectory hypotheses that minimizes the total cost, i.e., 
\begin{equation}
    \underset{A^{*}=(\mathbf{a}^1,...,\mathbf{a}^n)\in\mathfrak{A}}{\arg\min}\sum_{i\in\mathbb{X}}c(\mathbf{a}^i).
    \label{eq:globalminimize}
\end{equation}

The minimization problem (\ref{eq:globalminimize}) can be further posed as a multi-frame assignment problem by decomposing the constraint $A=(\mathbf{a}^1,...,\mathbf{a}^n)\in\mathfrak{A}$ into a set of smaller constraints, as first stated in \cite[Section III]{dualdecomposition}, in the form of
\begin{equation}
    \underset{\boldsymbol{\rho} \in \bigcup_{k=0}^{\tau} \mathcal{P}^k }{\arg\min}\sum_{i\in\mathbb{X}}\sum_{\mathbf{a}^i\in\mathcal{A}^i}c(\mathbf{a}^i)\rho(\mathbf{a}^i),
    \label{eq:costfunction}
\end{equation}
with the constraints sets denoted as
\begin{equation}
    \mathcal{P}^0 = \bigg \{\boldsymbol{\rho}\bigg|\sum_{\mathbf{a}^i\in\mathcal{A}^i}\rho(\mathbf{a}^i)=1,\forall i\in\mathbb{X}\bigg\},
    \label{eq:constraint1}
\end{equation}
\begin{multline}
    \mathcal{P}^k =\bigg \{\boldsymbol{\rho}\bigg|\sum_{i\in\mathbb{X}}\sum_{\substack{\mathbf{a}^i\in\mathcal{A}^i:\\(k,j)\in\mathcal{M}(\mathbf{a}^i)}}\rho(\mathbf{a}^i)=1,\forall j\in\mathbb{M}_k\bigg\},
    \label{eq:constraint2}
\end{multline}
where $k=1,...,\tau$, $\rho(\mathbf{a}^i)\in\{0,1\}$ is a binary indicator variable, indicating whether single trajectory hypothesis $\mathbf{a}^i$ in the $i$th track is included in a global hypothesis or not, and $\boldsymbol{\rho}=\{\rho(\mathbf{a}^i)\in\{0,1\}|\mathbf{a}^i\in\mathcal{A}^i~\forall ~i\in\mathbb{X}\}$ is the set of all binary indicator variables. The first constraint (\ref{eq:constraint1}) enforces that each global hypothesis should include one and only one single trajectory hypothesis from each track. The set of $\tau$ constraints (\ref{eq:constraint2}) enforce that each measurement from each scan should be associated to one and only one track.

\subsection{Multi-frame assignment via dual decomposition}


The multi-dimensional assignment problem (\ref{eq:costfunction}) is NP-hard for two or more scans of measurements. An effective solution to this is to use Lagrangian relaxation; this technique has been widely used to solve the multi-scan data association problem in the MHT framework \cite{lagrange1,lagrange2}. In this work, we focus on the dual decomposition formulation \cite{mrfdualdecomposition}, i.e., a special case of Lagrangian relaxation, whose competitive performance in solving the multi-frame assignment problem compared to traditional approaches \cite{lagrange1,lagrange2} has been demonstrated \cite{dualdecomposition}.

\subsubsection{Decomposition of Lagrangian dual}
We follow similar implementation steps as in \cite{dualdecomposition}. The original (primal) problem (\ref{eq:costfunction}) is separated into $\tau$ subproblems, one for each scan of measurements, and for each subproblem a binary variable $\boldsymbol{\rho}^k$ is used. The constraint that $\boldsymbol{\rho}^k$ are equal for each time scan $k$ is enforced through Lagrange multipliers that are incorporated into the subproblems to act as penalty weights. The $k$th subproblem can be written as \cite{dualdecomposition}
\begin{multline}
    \min_{\boldsymbol{\rho}^k\in\{\mathcal{P}^0,\mathcal{P}^k\}}\sum_{i\in\mathbb{X}}\sum_{\mathbf{a}^i\in\mathcal{A}^i}\bigg(\frac{c(\mathbf{a}^i)}{\tau}+\delta^k(\mathbf{a}^i)\bigg)\rho^k(\mathbf{a}^i) \\ \triangleq \min_{\boldsymbol{\rho}^k\in\{\mathcal{P}^0,\mathcal{P}^k\}}\mathcal{S}(\boldsymbol{\rho}^k,\boldsymbol{\delta}^k),
    \label{eq:subproblem}
\end{multline}
where the binary indicator variables and Lagrange multipliers used for the $k$th subproblem are denoted, respectively, by
\begin{subequations}
\begin{align}
    \boldsymbol{\rho}^k&=\{\rho^k(\mathbf{a}^i)\in\{0,1\}|\mathbf{a}^i\in\mathcal{A}^i~\forall~i\in\mathbb{X}\},\\
    \boldsymbol{\delta}^k&=\{\delta^k(\mathbf{a}^i)|\mathbf{a}^i\in\mathcal{A}^i~\forall~i\in\mathbb{X}\}.
\end{align}
\end{subequations}
The Lagrange multipliers $\delta^k(\mathbf{a}^i)$ can be any real number, but with the constraint that, for each single trajectory hypothesis, they must add up to zero over different subproblems  \cite{mrfdualdecomposition}. Thus, the set of Lagrange multipliers has the form
\begin{equation}
    \Lambda=\{\boldsymbol{\delta}^k|\sum_k\delta^k(\mathbf{a}^i)=0,~\forall\mathbf{a}^i\in\mathcal{A}^i~\forall~i\in\mathbb{X}\}.
\end{equation}

\subsubsection{Subproblem solving}
After eliminating all the constraints sets except two, i.e., $\mathcal{P}^0$ and $\mathcal{P}^k$, we obtain a 2-D assignment problem (\ref{eq:subproblem}). For the $k$th assignment problem, the objective is to associate each measurement received at time scan $k\leq\tau$, i.e., $j\in\mathbb{M}_k$, either to a pre-existing track or a new track at current time scan $\tau$, i.e., $i\in\mathbb{X}_{\tau}$, such that the total assignment cost is minimized. Problems of this type can be solved in polynomial time using a modified auction algorithm \cite[Chapter VII]{bar1990multitarget}. 

For a track that is created after time scan $k$, no measurement from time scan $k$ should be assigned to it; therefore, the measurement-to-track assignment cost is infinity. For a track that existed before and up to time scan $k$, i.e., $i\in\mathbb{X}_k$, if measurement $\mathbf{z}^j_k$ has never been associated to this track, let the measurement-to-track assignment cost be infinity; if otherwise, let the cost first be the minimum cost of the single trajectory hypotheses in this track that were updated by $\mathbf{z}^j_k$ \cite[Chapter VII, Equation (7.24)]{bar1990multitarget}, i.e.,
\begin{equation}
\min\sum_{\substack{\mathbf{a}^i\in\mathcal{A}^i:\\(k,j)\in\mathcal{M}(\mathbf{a}^i)}}\left(\frac{c(\mathbf{a}^i)}{\tau}+\delta^k(\mathbf{a}^i)\right).
\label{eq:minisinglecost}
\end{equation}

In order to keep the cost of a hypothesis that does not assign a measurement to a track the same for a pre-existing and a new track, the cost (\ref{eq:minisinglecost}) should then be subtracted by the minimum cost of hypotheses that this track is not updated by any of the measurements at time scan $k$, i.e.,
\begin{equation}
    \min\sum_{\substack{\mathbf{a}^i\in\mathcal{A}^i:\\(k,j)\notin\mathcal{M}(\mathbf{a}^i),\forall j\in\mathbb{M}_k}}\left(\frac{c(\mathbf{a}^i)}{\tau}+\delta^k(\mathbf{a}^i)\right).
\end{equation}
Note that, in the context of Lagrangian relaxation, the costs of single trajectory hypotheses refer to the costs that are penalized by the Lagrangian multipliers.

After solving this 2-D assignment problem, we can obtain the associations for each measurement at time scan $k$. For tracks not being associated to any measurements at time scan $k$, if the track is created before and up to time scan $k$, i.e., $i\in\mathbb{X}_k$, the single trajectory hypothesis
\begin{equation}
\argmin_{\mathbf{a}^i}\sum_{\substack{\mathbf{a}^i\in\mathcal{A}^i:\\(k,j)\notin\mathcal{M}(\mathbf{a}^i),\forall j\in\mathbb{M}_k}}\left(\frac{c(\mathbf{a}^i)}{\tau}+\delta^k(\mathbf{a}^i)\right)
\end{equation}
is included in the most likely global hypothesis; if otherwise, i.e., $i\in\mathbb{X}_{\tau}\setminus\mathbb{X}_k$, we can simply choose the single trajectory hypothesis
\begin{equation}
\argmin_{\mathbf{a}^i}\sum_{\mathbf{a}^i\in\mathcal{A}^i}\left(\frac{c(\mathbf{a}^i)}{\tau}+\delta^k(\mathbf{a}^i)\right)
\end{equation}
to be included in the most likely global hypothesis.

\subsubsection{Subgradient updates}
The objective of Lagrange relaxation is to find the tightest lower bound of the summation of the cost of each subproblem (\ref{eq:subproblem}). The dual problem can be expressed as \cite{dualdecomposition}
\begin{equation}
    \max_{\{\boldsymbol{\delta}^k\}\in\Lambda}\bigg(\sum_{k=1}^{\tau}\min_{\boldsymbol{\rho}^k\in\{\mathcal{P}^0,\mathcal{P}^k\}}\mathcal{S}(\boldsymbol{\rho}^k,\boldsymbol{\delta}^k)\bigg),
    \label{eq:dual}
\end{equation}
where the maximum can be found using subgradient methods \cite{subgradient}. The Lagrange multipliers $\{\boldsymbol{\delta}^k\}$ are updated  using
\begin{equation}
    \delta^k(\mathbf{a}^i) = \delta^k(\mathbf{a}^i) + \alpha_t\cdot g^k(\mathbf{a}^i),
\end{equation}
where $g^k(\mathbf{a}^i)$ is the projected subgradient that can be calculated as
\begin{equation}
    g^k(\mathbf{a}^i) = \rho^k(\mathbf{a}^i)-\frac{1}{\tau}\sum_{k^{\prime}=1}^{\tau}\rho^{k^{\prime}}(\mathbf{a}^i),
    \label{eq:constraintLagrange}
\end{equation}
and $\alpha_t$ is the step size at iteration $t$. There are many rules to set the step size, see \cite{mrfdualdecomposition}. In this work, we choose to use the same setting as in \cite{dualdecomposition}, which has the form
\begin{equation}
    \alpha_t = \frac{\text{BESTPRIMAL}_t-\text{DUAL}_t}{\|\{\mathbf{g}^k\}\|^2},
\end{equation}
where $\text{BESTPRIMAL}_t$ is the best (minimum) feasible primal cost so far obtained, and $\text{DUAL}_t$ is the dual cost calculated at iteration $t$ from (\ref{eq:dual}). The optimal solution is assumed to be attained when the relative gap between the primal cost and the dual cost is less than a specified threshold $\epsilon$ \cite{mrfdualdecomposition}, where
\begin{equation}
    \text{gap} = \frac{\text{BESTPRIMAL}_t-\text{DUAL}_t}{\text{BESTPRIMAL}_t}.
\end{equation}

Each subproblem solution will, in general, be infeasible with respect to the primal problem (\ref{eq:costfunction}); nevertheless, subproblem solutions will usually be nearly feasible since large constraints violations got penalized \cite{mrfdualdecomposition}. Hence, feasible solutions $\boldsymbol{\rho}$ can be obtained by correcting the minor conflicting binary elements on which subproblem solutions $\boldsymbol{\rho}^k$ disagree. For those tracks that we have not yet selected which single trajectory hypothesis to included in the most likely global hypothesis, we use the branch and bound technique to reconstruct the best feasible solution at each iteration of the Lagrange relaxation. Note that there are many other ways to recover a feasible primal solution from subproblem solutions, see \cite{mrfdualdecomposition}.

\subsection{Discussion}

As discussed in \cite{pmbmpoint}, the structure of the data association hypotheses in the PMBM density is similar to that of the track-oriented MHT. Benefiting from the separability of the global hypothesis weight (\ref{eq:globalhypothesisweight}), in this work, we make use of the same approximation technique in the track-oriented MHT to find the most likely global hypothesis by solving a multi-frame assignment problem. It should be noted that, although the obtained, most likely global hypothesis is an MB, the overall approximated filtering density
describing the set of detected trajectories, implicitly represented, is an MBM. The objective of solving the multi-frame assignment problem is to know which Bernoulli components are included in the MB with the highest weight. At each update step, Bernoulli components forming the highest weight MB, together with the other Bernoulli components that survive from $N$-scan pruning, are carried over to the next filtering recursion. Because the data association ambiguity is assumed to be resolved before time $\tau-N$, obtaining the most likely global hypothesis at time $\tau$, which explains the origin of each measurement from scan $\tau-N$ to current scan $\tau$, requires the solution of a $N+2$ dimensional assignment problem \cite{blackman2004multiple}.


In contrast, the existing PMBM target filter in \cite{pmbmpoint} operates by propagating a number of global hypotheses with explicitly calculated weights over time. The number of global hypotheses grows exponentially over time due to data association; thus it is intractable to enumerate all the possible global hypotheses. In order to have a tractable solution, each predicted global hypothesis only generates a set of most likely updated global hypotheses by approximately selecting the $M$ most likely data associations, using, e,g., Murty's algorithm \cite{murty} or Gibbs sampling \cite{gibbsglmb}. Further, updated global hypotheses with negligible normalized weights are assumed to be unlikely to generate any high-weight global hypothesis in subsequent time steps; thus they can be pruned. Only maintaining global hypotheses with non-negligible weights means that we prune single target hypotheses that are not included in any of the remaining global hypotheses. One can relate this pruning procedure to the N-scan pruning used in the PMBM trajectory filter, in which the same intuition applies: single trajectory hypotheses are pruned if their updates are assumed to be unlikely to be included in the most likely global hypothesis in subsequent time steps.

The computational complexity of both filters can be further reduced by limiting the number of single target/trajectory hypotheses. For the PMBM target filter, we can prune single target hypothesis with small enough Bernoulli existence probability. A more efficient approximation method is called recycling \cite{recycle}, in which Bernoulli components are first approximated as being Poisson and then added to the PPP density representing undetected targets. The benefits of applying recycling in the PMBM target filter has been demonstrated in \cite{performanceevaluation}. As for the PMBM trajectory filter, pruning single trajectory hypotheses will sometimes harm the solvability of the multi-frame assignment problem, since the problem is formulated using the measurement assignment information contained in single trajectory hypotheses. Instead, we can choose single trajectory hypotheses $\mathbf{a}^i\in\mathcal{A}^i, \forall i\in\mathbb{X}$ with small Bernoulli existence probability $r^{\mathbf{a}^i}$ at time step $k$ to be updated only by missed detection at time step $k+1$. Then single trajectory hypotheses with several consecutive missed detections can be pruned.

\section{Simulations and Results}
In this section, we show simulation results that compare the proposed PMBM trajectory filter with the PMBM target filter \cite{pmbmpoint2}, and the $\delta$-GLMB filter with joint prediction and update steps \cite{gibbsglmb}. For the $\delta$-GLMB filter and the PMBM target filter, the $M$ most likely data associations are approximately found using the Gibbs sampling solution proposed in \cite{gibbsglmb}. In the implementation, all the codes are written in \texttt{MATLAB}, except the Gibbs sampling and the auction algorithm, which are written in \texttt{C++}. 


\subsection{State space model}
A two-dimensional Cartesian coordinate system is used to define measurement and target kinematic parameters. The kinematic target state is a vector of position and velocity $x_k=[p_{x,k},p_{y,k},\dot{p}_{x,k},\dot{p}_{y,k}]^T$. A single measurement is a vector of position $z_k=[z_{x,k},z_{y,k}]^T$. Targets follow a linear Gaussian constant velocity model $f_{k|k-1}(x_k|x_{k-1}) = \mathcal{N}(x_k;F_kx_{k-1},Q_k)$, with parameters
\begin{equation*}
    F_k=I_2 \otimes \begin{bmatrix}
        1 & T\\
        0 & 1
    \end{bmatrix}, \quad Q_k =  0.002I_2\otimes\begin{bmatrix}
        T^3/3 & T^2/2\\
        T^2/2 & T
    \end{bmatrix},
\end{equation*}
where $\otimes$ is the Kronecker product, $I_m$ is an identity matrix of size $m\times m$, and $T=1$. The linear Gaussian measurement likelihood model has density $g(z_k|x_k) = \mathcal{N}(z_k;H_kx_k,R_k)$, with parameters $H_k = I_2\otimes[1,0]$ and $R_k = I_2$. 

\subsection{Parameter setting}
\begin{figure}[!t]
    \centering
    \includegraphics[width=0.35\textwidth]{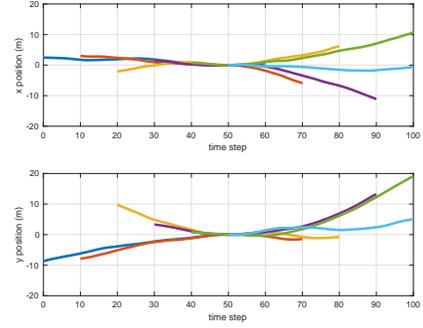}
    \caption{Scenario of simulations. There are six targets born at times \{1,11,21,31,41,51\} and dead at times \{61,71,81,91,101,101\} respectively. Targets move from left to right, and they are in close proximity around the mid-point.}
    \label{fig:coal}
\end{figure}

We consider $101$ time steps and the scenario in  Figure \ref{fig:coal}. For each trajectory in ground truth, we initiate the midpoint from a Gaussian with mean $[0,0,0,0]^T$ and covariance matrix $10^{-6}I_4$, and the rest of the trajectory is generated by running forward and backwards dynamics. This scenario is challenging due to the high number of targets in close proximity and the fact that one target is born when and where the other five are in close proximity. In the simulation, we consider cases with constant target survival probability $p^s=0.99$, constant detection probability $p^d\in\{0.7,0.9\}$, and Poisson clutter uniform in the region of interest with rate $\lambda^c\in\{10,30\}$.

For the PMBM target/trajectory filter, targets are born according to a PPP with intensity $0.05$ and Gaussian density with mean $[0,0,0,0]^T$ and covariance matrix $\texttt{diag}([100^2,1,100^2,1])$, which covers the region of interest. For the PPP part, components are pruned with weight smaller than $10^{-4}$. For the PMBM target filter, Bernoulli components with existence probability less than $0.1$ are recycled. For the $\delta$-GLMB filter, targets are born according to a Bernoulli process with existence probability $0.05$ and the same Gaussian density as the PPP birth model. For the PMBM target filter and the $\delta$-GLMB filter, for a global hypothesis with weight $\mathcal{W}$, we use Gibbs sampling to select at most $\lceil \mathcal{W}\cdot N_h \rceil$ updated global hypotheses with the highest weights. In addition, global hypotheses are truncated to only contain those that correspond to $99.99\%$ of the likelihood. For all the filters, gating is performed to remove unlikely target-to-measurement association. 

An implementation of the PMBM trajectory filter propagating the whole trajectory (full state sequence) using information Gaussian representation has been provided in \cite{continuityPMBM}. A more computational efficient implementation is to consider only the target states of the last $L$ steps \cite{garcia2016trajectory}. In the simulation, we compare three filters: the $\delta$-GLMB filter, which provides filtered estimates; a PMBM trajectory filter that provides filtered estimates ($L=1$); and a PMBM trajectory filter that provides smoothed estimates ($L$ is equal to trajectory length). Smoothed estimates can be easily obtained using Rauch-Tung-Striebel smoother, since the full target state sequence is contained in single trajectory hypotheses. As for the $\delta$-GLMB filter, a closed form for forward-backward smoothing was developed in \cite{glmbsmoother} but a tractable implementation is not yet provided.


\subsection{Performance evaluation}
Given a multi-target posterior density, several state estimators are available \cite{glmbpoint}. In this work, we choose to extract the target states by finding the maximum a posteriori cardinality estimate, following the methods suggested in \cite{glmbpoint,pmbmpoint2}. Filtering performance is assessed using the generalized optimal subpattern (GOSPA) metric \cite{gospa} ($\alpha=2,c=20,p=1$), which allows for decomposing the estimation error into location error, missed detection error and false detection error. 

The $\delta$-GLMB filter provides estimates of a target trajectory by estimating the current state at each time, and connecting estimates from different times with the same label. For the PMBM trajectory filter, we can choose to either provide the trajectory estimation directly from single trajectory hypothesis density \cite{continuityPMBM} or connect current state estimates which belong to the same track. In the simulation, the latter one is used, and we only score the final trajectory estimates\footnote{The quality of the trajectory estimates can also be evaluated in an online fashion. For instance, we can compute the trajectory estimation error at each time instant and normalize the error by the trajectory length, see \cite{garcia2016trajectory}.}. For assessing the estimation performance on the sets of trajectories for the $\delta$-GLMB filter and the PMBM trajectory filter, we use the metric proposed in \cite{trajectorymetric} ($c=20,p=1,\gamma=2$), which allows for decomposing the estimation error into location error, missed detection error, false detection error and error due to track switches.

\subsection{Results}
\begin{figure*}[!t]
    \centering
    \includegraphics[width=0.25\textwidth]{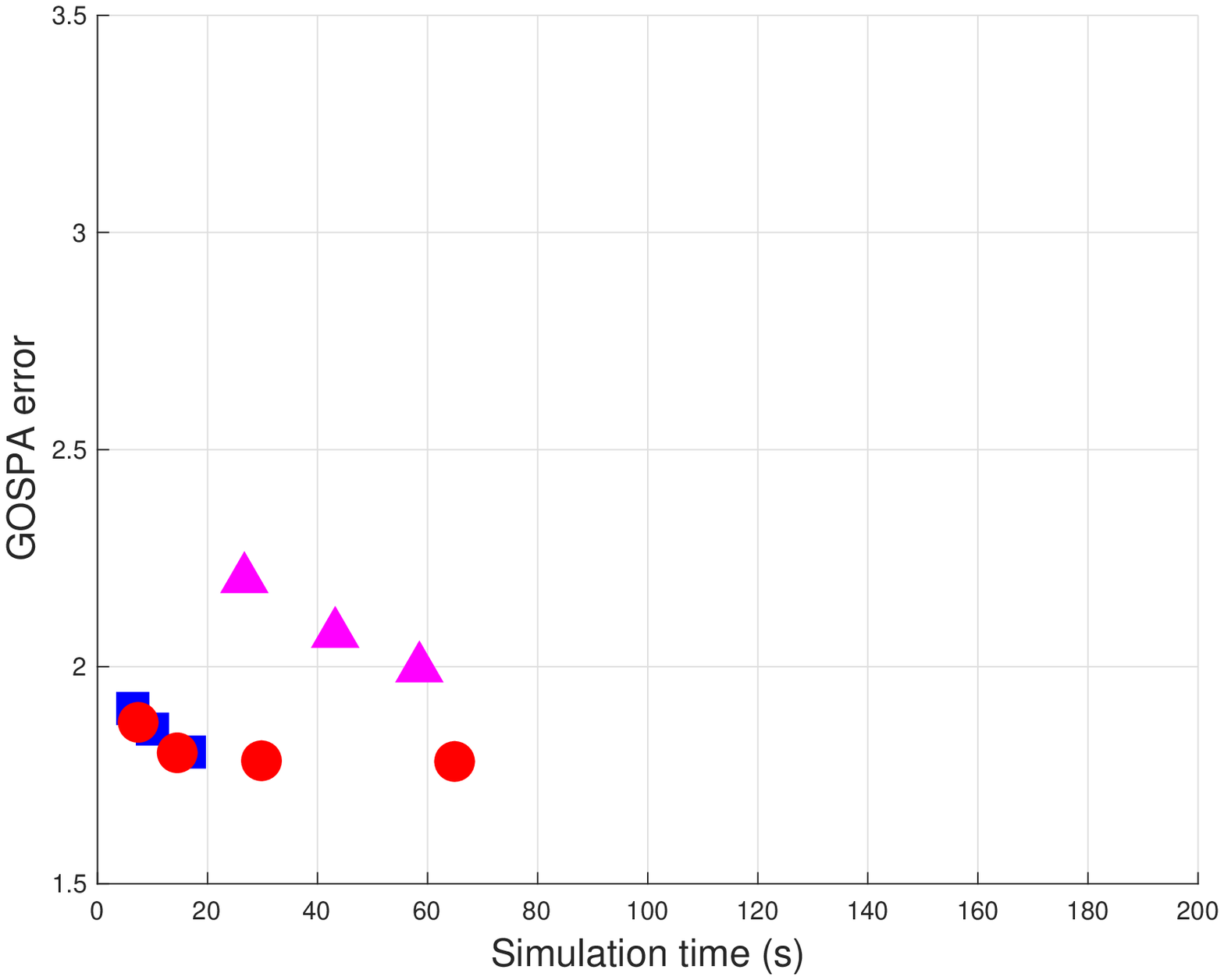}
    \hspace{-3ex}
    \includegraphics[width=0.25\textwidth]{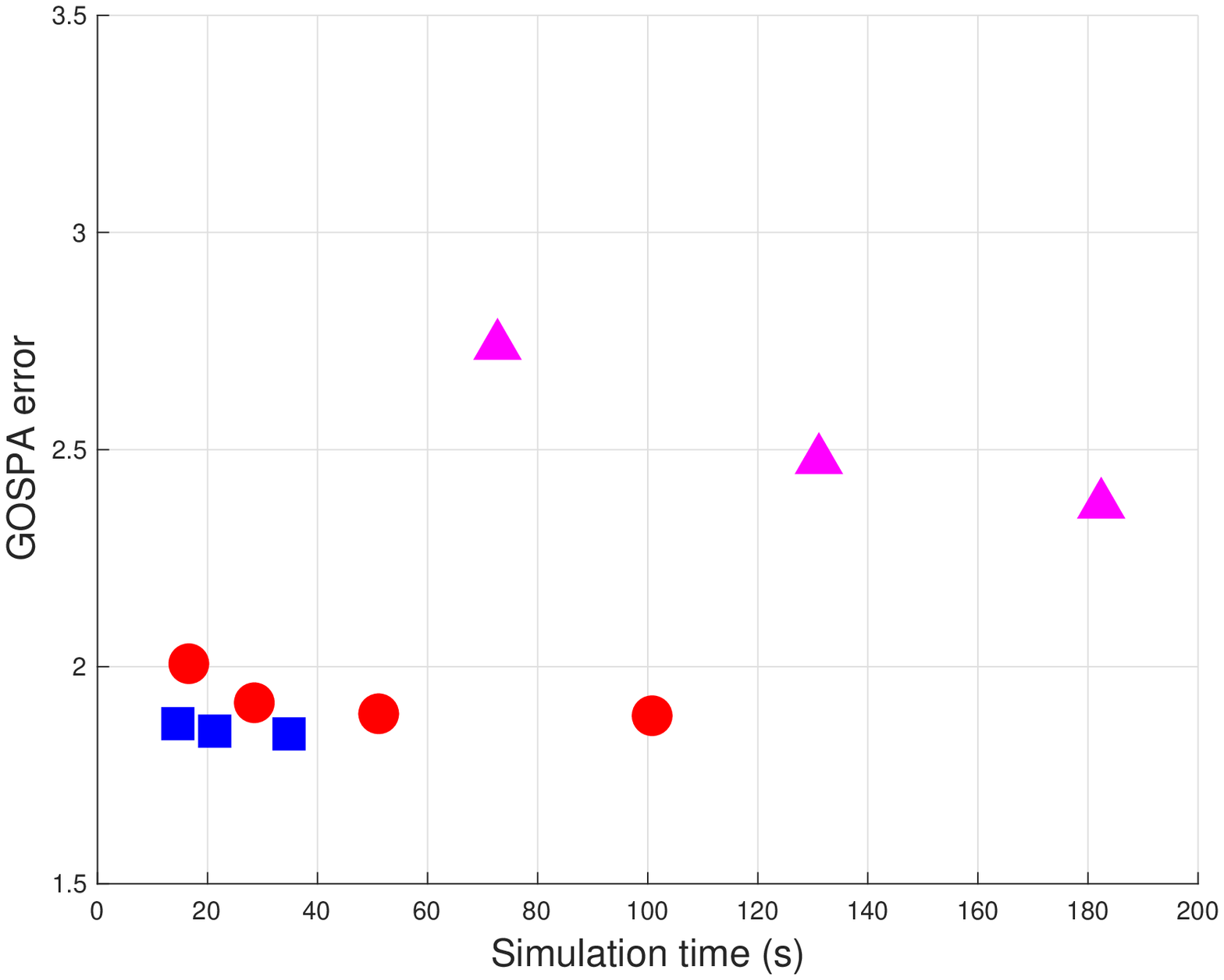}
    \hspace{-3ex}
    \includegraphics[width=0.25\textwidth]{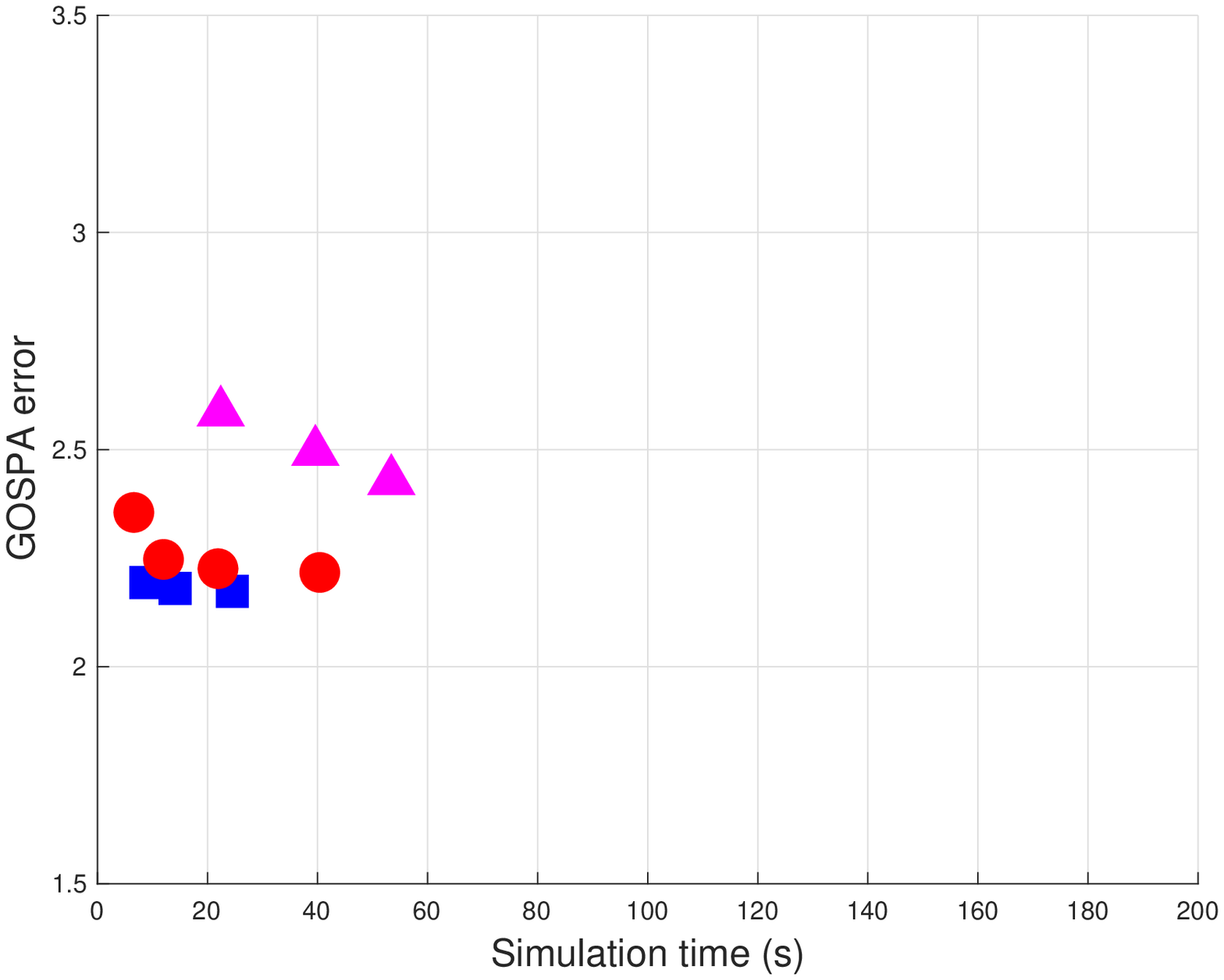}
    \hspace{-3ex}
    \includegraphics[width=0.25\textwidth]{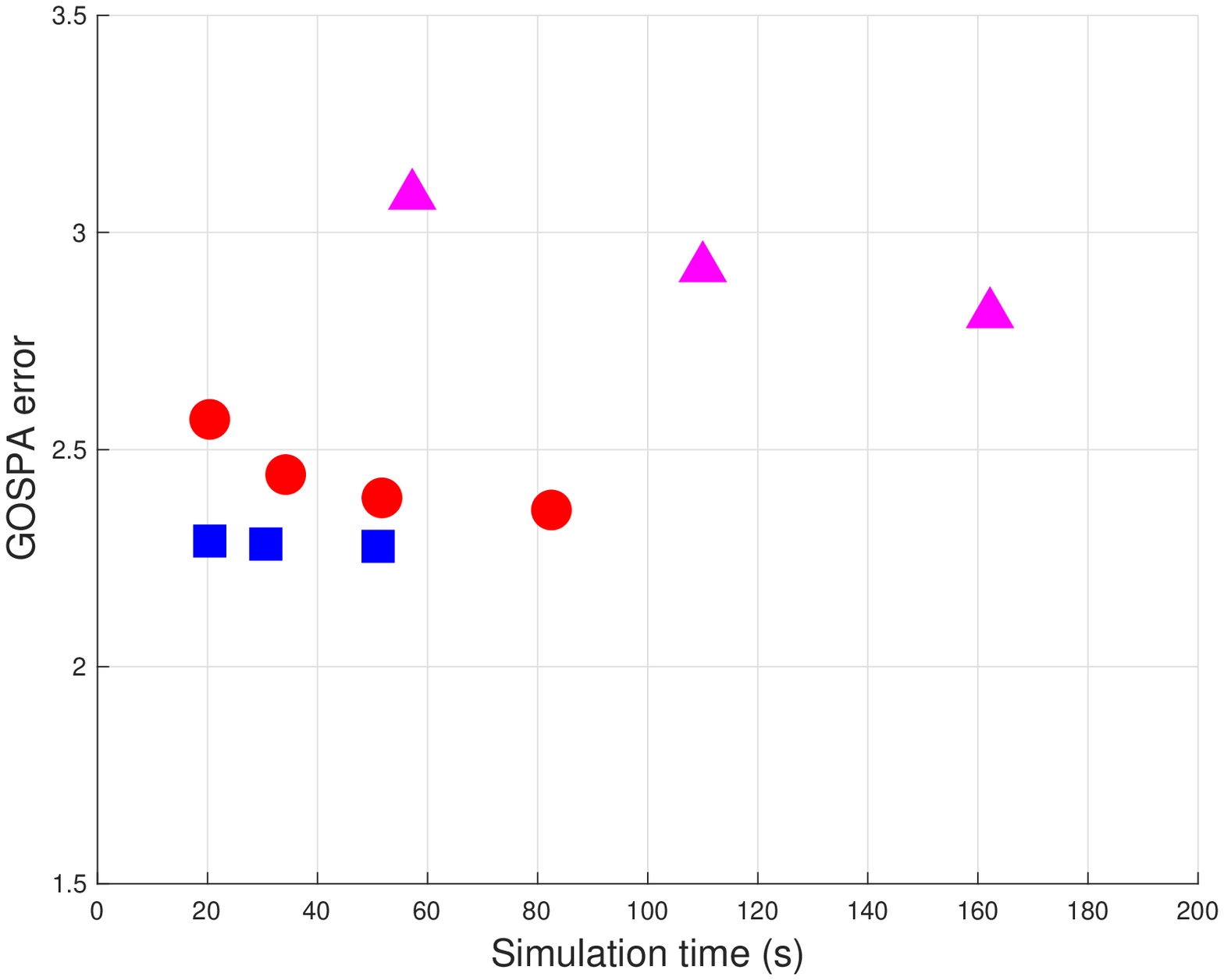}
    \caption{Performance comparison among the $\delta$-GLMB filter (triangle), the PMBM target filter (rectangle) and the PMBM trajectory filter (circular): simulation time versus GOSPA error (in logarithm). Plots from left to right, respectively, correspond to scenario parameter setting: 1) $p^d=0.9,\lambda^c=10$, 2) $p^d=0.9,\lambda^c=30$, 3) $p^d=0.7,\lambda^c=10$ and 4) $p^d=0.7,\lambda^c=30$. For the same filter in each subplot, the scatter points from left to right marked with the same color, respectively, correspond to $N_h=\{100,200,300\}$ for the $\delta$-GLMB filter, $N_h=\{50,100,200\}$ for the PMBM target filter, and $N=\{2,3,4,5\}$ for the PMBM trajectory filter.}
    \label{fig:scatterplot}
\end{figure*}

The results are obtained over 100 Monte Carlo trials. The average localization error per target, missed and false detection error, measured using GOSPA, of different filters are listed in Table \ref{tab:SimulationResults}. The performance comparison regarding computational time and GOSPA error among different filters is presented in Fig. \ref{fig:scatterplot}. Overall, the PMBM target filter has the best filtering performance, followed by the PMBM trajectory filter, with the $\delta$-GLMB filter in last. 




Table \ref{tab:trajectoryEstimation} presents the trajectory estimation error\footnote{We compared all the trajectories that had existed in the simulation to the ground truth (six true trajectories).} of the $\delta$-GLMB filter and the PMBM trajectory filter.  By only looking at the trajectory estimation results without smoothing, the PMBM trajectory filter has significantly smaller missed detection error, similar track switch error, and slightly larger localization and false detection error than the $\delta$-GLMB filter. In the simulation, we found that, for the $\delta$-GLMB, missed detection frequently occurs when targets become in close proximity, which leads to the result that the trajectories obtained using the $\delta$-GLMB filter are usually shorter than those obtained by the PMBM filter. This explains the results since a shorter trajectory implies lower error due to track switches, localization mismatch and possible false detections.


\begin{table*}[!t]
\caption{Simulation results: the average state estimation error per time step. Legend: te-- total gospa error; le--average location error per target; mt--missed targets; ft--false targets; $\textsc{n}_h$: the maximum number of updated global hypotheses per predicted global hypothesis; $\textsc{n}$: n-scan pruning; glmb: $\delta$-glmb filter with joint prediction and update steps; pmmt: pmbm target filter; pmdd: pmbm trajectory filter}
\label{tab:SimulationResults}
\centering
\resizebox{\textwidth}{!}{%
\begin{tabular}{c | cccc | cccc | cccc | cccc }
& \multicolumn{4}{c|}{$p^d=0.9, \lambda^c=10$} & \multicolumn{4}{c|}{$p^d=0.9, \lambda^c=30$} & \multicolumn{4}{c|}{$p^d=0.7, \lambda^c=10$} & \multicolumn{4}{c}{$p^d=0.7, \lambda^c=30$}\\
Filter &  \textsc{ te } &  \textsc{ le } & \textsc{ mt } & \textsc{ ft } &  \textsc{ te }& \textsc{ le } & \textsc{ mt } & \textsc{ ft } &  \textsc{ te }& \textsc{ le } & \textsc{ mt } & \textsc{ ft } &  \textsc{ te }& \textsc{ le } & \textsc{ mt } & \textsc{ ft }  \\
\hline
\textsc{glmb}($\textsc{n}_h=300$)                           &$7.36$  & $0.81$ & $4.40$ & $0.34$ &$10.73$& $0.77$ & $8.12$ & $0.36$ &$11.33$& $0.86$ & $8.27$ & $0.61$ &$16.65$& $0.82$ & $14.21$ & $0.49$   \\
\textsc{glmb}($\textsc{n}_h=200$)                            &$7.97$ & $0.80$ & $5.08$ & $0.34$ &$11.89$& $0.76$ & $9.39$ & $0.35$ &$12.12$& $0.86$  & $9.16$ & $0.57$ &$18.50$& $0.79$ & $16.36$  & $0.42$ \\
\textsc{glmb}($\textsc{n}_h=100$)                            &$9.04$ & $0.79$ & $6.21$ & $0.38$ &$15.47$& $0.74$ & $13.35$ & $0.28$ &$13.27$& $0.83$  & $10.37$ & $0.62$ &$21.86$& $0.76$ & $20.16$  & $0.31$  \\

\hline
\textsc{pm}\textsc{mt}($\textsc{n}_h=200$)                        &$6.07$ & $0.79$ & $2.75$ & $0.52$ &$6.33$& $0.80$ & $3.02$ & $0.54$ &$8.79$& $0.88$  & $5.19$ & $0.78$&$9.75$ & $0.87$ & $6.19$  & $0.82$  \\
\textsc{pm}\textsc{mt}($\textsc{n}_h=100$)                        &$6.40$ & $0.76$ & $3.12$ & $0.53$ &$6.37$& $0.79$ & $3.06$ & $0.55$  &$8.85$& $0.88$  & $5.24$ & $0.79$ &$9.80$& $0.88$ & $6.24$  & $0.80$   \\
\textsc{pm}\textsc{mt}($\textsc{n}_h=50$)                        &$6.71$ & $0.74$ & $3.45$ & $0.54$ &$6.48$& $0.78$ & $3.19$ & $0.54$ &$8.97$&$0.86$  & $5.38$ & $0.78$ &$9.87$& $0.87$ & $6.32$  & $0.81$   \\

\hline
\textsc{pm}\textsc{dd}($\textsc{n}=5$)                                                  &$5.94$ & $0.83$ & $2.63$ & $0.42$ &$6.60$& $0.82$ & $3.35$ & $0.45$&$9.18$& $0.90$  & $5.57$ & $0.78$ &$10.65$& $0.89$ & $7.20$  & $0.78$  \\
\textsc{pm}\textsc{dd}($\textsc{n}=4$)                                                  &$5.98$ & $0.83$ & $2.69$ & $0.40$ &$6.63$& $0.82$ & $3.41$ & $0.44$ &$9.26$& $0.90$  & $5.67$ & $0.76$ &$10.91$& $0.89$ & $7.52$  & $0.73$  \\
\textsc{pm}\textsc{dd}($\textsc{n}=3$)                                                  &$6.06$ & $0.83$ & $2.81$ & $0.39$ &$6.80$& $0.82$ & $3.62$ & $0.42$ &$9.46$& $0.91$  & $5.91$ & $0.72$&$11.51$ & $0.89$ & $8.28$  & $0.64$\\
\textsc{pm}\textsc{dd}($\textsc{n}=2$) &$6.50$ & $0.82$ & $3.41$ & $0.34$ &$7.44$& $0.81$ & $4.39$ & $0.38$&$10.54$ & $0.91$  & $7.25$ & $0.61$ &$13.06$& $0.89$ & $10.10$  & $0.53$\\
\end{tabular}
}
\end{table*}

\begin{table*}[!t]
\caption{Simulation results: the sum of trajectory estimation error. Legend: te--total error; le--location error; me--missed detection; fe--false detection; ts--track switch error; pmdds: pmbm trajectory filter with smoothing}
\label{tab:trajectoryEstimation}
\centering
\resizebox{\textwidth}{!}{%
\begin{tabular}{c | ccccc | ccccc | ccccc | ccccc }
& \multicolumn{5}{c|}{$p^d=0.9, \lambda^c=10$} & \multicolumn{5}{c|}{$p^d=0.9, \lambda^c=30$} & \multicolumn{5}{c|}{$p^d=0.7, \lambda^c=10$} & \multicolumn{5}{c}{$p^d=0.7, \lambda^c=30$}\\
Filter & \textsc{ te } & \textsc{ le } & \textsc{ me } & \textsc{ fe }& \textsc{ ts } & \textsc{ te } &\textsc{ le } & \textsc{ me } & \textsc{ fe }& \textsc{ ts } &\textsc{ te } & \textsc{ le } & \textsc{ me } & \textsc{ fe }& \textsc{ ts } & \textsc{ te } &\textsc{ le } & \textsc{ me } & \textsc{ fe }& \textsc{ ts }  \\
\hline
\textsc{glmb}($\textsc{n}_h=300$)                           & $857$  & $348$ & $440$ & $34$ & $35$ & $1176$ & $299$ & $812$ & $36$ & $29$ & $1242$ & $327$ & $826$ & $60$ & $29$ & $1759$& $269$  & $1420$ & $42$ & $28$\\
\textsc{glmb}($\textsc{n}_h=200$)                            & $918$  & $340$ & $508$ & $35$ & $35$ & $1287$& $287$ & $938$ & $35$ & $27$ & $1319$& $317$ & $915$  & $56$ & $31$ & $1927$& $241$ & $1628$ & $34$  & $24$\\
\textsc{glmb}($\textsc{n}_h=100$)                            & $1021$  & $329$ & $621$ & $37$ & $34$ & $1631$& $247$ & $1333$ & $27$ & $24$ & $1434$& $304$  & $1040$ & $61$ & $29$ & $2324$& $195$ & $2010$ & $24$  & $95$\\

\hline
\textsc{pm}\textsc{dd}($\textsc{n}=5$)            & $717$& $379$ & $262$ & $42$ & $34$ & $778$& $366$ & $335$ & $45$ & $32$ & $1032$& $370$  & $556$ & $77$ & $29$ & $1174$& $353$ & $720$  & $74$ & $27$\\
\textsc{pm}\textsc{dd}($\textsc{n}=4$)            & $714$& $375$ & $268$ & $39$ & $32$ & $777$& $362$ & $341$ & $44$ & $30$ & $1039$& $370$ & $566$  & $76$ & $27$ & $1198$& $350$ & $750$ & $71$  & $27$\\
\textsc{pm}\textsc{dd}($\textsc{n}=3$)           & $727$ & $375$ & $281$ & $38$ & $33$ & $795$& $360$ & $362$ & $42$ & $31$ & $1057$& $371$ & $589$  & $70$ & $27$ & $1257$& $343$ & $826$ & $62$  & $26$\\
\textsc{pm}\textsc{dd}($\textsc{n}=2$)          & $769$  & $360$ & $341$ & $34$ & $34$ & $857$& $348$ & $439$ & $38$ & $32$ & $1163$& $354$ & $722$  & $58$ & $29$ & $1405$& $326$ & $1005$ & $48$  & $26$\\

\hline
\textsc{pm}\textsc{dds}($\textsc{n}=5$)           & $420$ & $263$ & $123$ & $14$ & $20$ & $472$& $258$ & $181$ & $14$ & $19$ & $701$& $256$  & $411$ & $16$ & $18$ & $842$& $245$ & $561$  & $18$ & $18$\\
\textsc{pm}\textsc{dds}($\textsc{n}=4$)           & $416$ & $261$ & $127$ & $9$ & $19$ & $471$& $256$ & $185$ & $12$ & $18$ & $720$& $255$ & $430$  & $17$ & $18$ & $883$& $241$ & $609$ & $16$  & $17$\\
\textsc{pm}\textsc{dds}($\textsc{n}=3$)            & $453$& $257$ & $170$ & $7$ & $19$ & $514$& $250$ & $234$ & $12$ & $18$ & $766$& $255$ & $478$  & $14$ & $19$ & $976$& $233$ & $710$ & $15$  & $18$\\
\textsc{pm}\textsc{dds}($\textsc{n}=2$)           & $529$ & $245$ & $259$ & $6$ & $19$ & $618$& $241$ & $348$ & $11$ & $18$ & $926$& $241$ & $654$  & $11$ & $20$ & $1177$& $220$ & $925$ & $13$ &$19$\\
\end{tabular}
}
\end{table*}

\section{Conclusion}
This paper has proposed an efficient implementation of the PMBM trajectory filter. Compared with the existing PMBM target filter, the PMBM trajectory filter is able to estimate all target trajectories. Also, a simulation study shows that the PMBM trajectory filter has better performance than the $\delta$-GLMB filter in terms of state/trajectory estimation error and computational time in a scenario with coalescence.



%





\ifCLASSOPTIONcaptionsoff
  \newpage
\fi



%


\bibliographystyle{IEEEtran}
\bibliography{IEEEabrv,mybibli}

%








\end{document}